\documentclass[11pt,a4paper]{article}
\pdfoutput=1

\usepackage{bm}
\usepackage{amsmath,amssymb}
\usepackage{mathtools}
\usepackage{cite}

\usepackage{tikz, tcolorbox}
\usepackage{graphicx}
\usepackage{subcaption}
\usepackage{color}
\usepackage{upgreek}

\usepackage{hyperref} 
\hypersetup{colorlinks=true, citecolor=blue, filecolor=black, linkcolor=blue,urlcolor=blue, pdfpagemode=UseNone}

\numberwithin{figure}{section}
\numberwithin{equation}{section}

\newcommand{\nablabar}{\overline{\nabla}}
\newcommand{\be}{\begin{equation}}
\newcommand{\ee}{\end{equation}}
\newcommand{\bea}{\begin{eqnarray}}
\newcommand{\eea}{\end{eqnarray}}
\def\beal#1\eeal{\begin{align}#1\end{align}}   
\def\besp#1\eesp{\begin{multline}#1\end{multline}} 

\newcommand\ie{\textit{i.e.}\ }
\newcommand\eg{\textit{e.g.}\ }

\makeatletter
\providecommand*{\shuffle}{%
  \mathbin{\mathpalette\shuffle@{}}%
}
\newcommand*{\shuffle@}[2]{%
  \sbox0{$#1\vcenter{}$}%
  \kern .15\ht0 
  \rlap{\vrule height .25\ht0 depth 0pt width 2.5\ht0}%
  \raise.1\ht0\hbox to 2.5\ht0{%
    \vrule height 1.75\ht0 depth -.1\ht0 width .17\ht0 %
    \hfill
    \vrule height 1.75\ht0 depth -.1\ht0 width .17\ht0 %
    \hfill
    \vrule height 1.75\ht0 depth -.1\ht0 width .17\ht0 %
  }%
  \kern .15\ht0 
}
\makeatother

\newcommand{\hh}{\mathcal{H}}

\usepackage[a4paper,hmarginratio=1:1,vmarginratio=2:3,totalwidth=15.6cm,totalheight=22.cm]{geometry}

\long\def\symbolfootnote[#1]#2{\begingroup
\def\thefootnote{\fnsymbol{footnote}}\footnote[#1]{#2}\endgroup}

\begin{document}
\thispagestyle{empty}

\begin{flushright}
  \today
\end{flushright}
\vspace{3.5cm}

\begin{center}
  {\Large \bf Weyl gauge symmetry at LIGO-Virgo-KAGRA}
\end{center}
\vskip1cm

\begin{center}
{{\bf D. M. Ghilencea}
 \symbolfootnote[1]{E-mail: dumitru.ghilencea@cern.ch}
 and
 {\bf V.-M.  Mandric}
 \symbolfootnote[2]{E-mail: vlad.mandric@theory.nipne.ro}}
\end{center}

\begin{center}{
{\it Department of Theoretical Physics, \\ National Institute of Physics
and Nuclear Engineering (IFIN), \\ Bucharest 077125, Romania}}\\
\date{\today}
\end{center}

\begin{abstract}
With current advances in gravitational wave (GW) detection made by the worldwide
LIGO-Virgo-KAGRA (LVK) network of detectors, ever-more sensitive tests of gravity in
the strong-field regime are now possible.
This enables one to test  gauge theories beyond Einstein-Hilbert action,
such as Weyl gauge theories of gravity. The only anomaly-free (quantum) gauge theory of
a space-time symmetry beyond Poincar\'e is   based on  Weyl gauge
group (of  dilatations and Poincar\'e symmetry) with Weyl conformal geometry as
its natural underlying geometry. This gauge theory has spontaneous
breaking of Weyl gauge symmetry to Einstein-Hilbert and Proca actions,
plus a positive cosmological constant.  We investigate  the GW polarisation
modes of Weyl (quadratic) gauge theory of gravity in Weyl geometry and
compare our findings to the most recent experimental data. We show how the geodesic deviation
equation from Riemannian geometry translates  to Weyl geometry, and explain
why it is crucial to perform the analysis around de Sitter
background, which is the correct low-energy limit of Weyl  quadratic gravity,
to not alter the GW content, and then compute the polarisation modes. In addition to
the two transverse-traceless tensor modes predicted by Einstein-Hilbert action,
we find two additional vector  modes induced by the transverse fluctuations
of the Weyl  gauge field. If detected, these vectors modes would be important evidence for Weyl gauge symmetry.
\end{abstract}

\newpage

\tableofcontents

\section{Introduction}

The Einstein-Hilbert action represents one of the crowning
achievements of modern physics. It has been thoroughly tested and very successful in
explaining phenomena across a wide range of scales. However, it comes short of being a truly
fundamental theory. For example, the presence of spacetime singularities
\cite{PhysRevLett.14.57,Hawking:1970zqf} or the failure of the Einstein-Hilbert (EH) action to
be (perturbatively) renormalisable in the usual way \cite{tHooft:1974toh,Goroff:1985th}
highlight some of the limitations of EH gravity as a quantum theory. These
problems,
together with the adage that \textit{gravity ``is'' geometry} taken at face value, suggest that a
potential way forward in constructing a (quantum) gauge theory of gravity is to go beyond
Riemannian geometry.

At high energies, such as those present in the early Universe, all physical
states can be regarded as effectively massless. This may indicate that both gravity
and the Standard Model (SM) could emerge from a more fundamental theory endowed with a form
of scale symmetry (global, local or gauged) within which all mass scales, including the
Planck scale, are generated dynamically by spontaneous breaking of this symmetry via
vacuum expectation values (vev) of some scalar fields. Since it may actually be fundamental, we
choose a {\it gauged} scale invariance. It is well-known that the only \textit{true} gauge theory
with a space-time symmetry beyond Poincar\'e symmetry,
\ie with a dynamical (physical) gauge boson, is  the gauge theory of Weyl
group (of dilatations and Poincar\'e symmetry) \cite{Condeescu:2024cjh,CDA}. The natural underlying geometry for this symmetry is  Weyl conformal
geometry \cite{Weyl1,Weyl2,Weyl3}, which has this symmetry 
{\it by definition}. Gauging the algebra of the Weyl group \cite{Tait}
leads one to Weyl (quadratic) gauge theory of gravity (WQG) action
\cite{Condeescu:2024cjh,CDA}
as introduced by Weyl more than a century ago \cite{Weyl2}.
This is a \textit{vector-tensor} gauge theory of gravity,
while  Weyl geometry is, naturally, just a  ``covariantised'' version of Riemannian geometry
with respect to the (additional) gauged dilatation symmetry.

Historically,
Weyl initially thought that WQG  would describe gravity and electromagnetism, but it obviously
failed, we now know that electromagnetism corresponds to an internal U(1) gauge symmetry not to 
external (space-time) real dilatations symmetry. But at that time Einstein argued  that due to the so-called non-metricity of the
geometry (\ie $\tilde \nabla_\mu g_{\alpha\beta} \neq 0$) there is a \textit{second clock effect}
\cite{Weyl1} which  contradicts all experimental evidence. This argument lasted a century
and outlived most attempts to restore Weyl geometry as a physical theory (of gravity),
with notable exceptions of Dirac \cite{Dirac} and Smolin \cite{Smolin}. They
saw the importance of Weyl gauge symmetry and constructed such simplified theories linear in curvature
(for a  brief review see \cite{Fall,review};
for a historical review and references see \cite{Scholz}).

The problem with Einstein's argument is that it overlooked the fact that in the Weyl
symmetric phase the action has {\it no mass scale}, and without a mass there is no
clock rate, therefore no \textit{second clock effect}.
Moreover, in the Weyl gauge {\it covariant formulation} of Weyl geometry,
required in a gauge theory, the theory is actually {\it metric,} $\hat \nabla_\mu g_{\alpha\beta}=0$,
something known since Dirac \cite{Dirac} but overlooked until
very recently \cite{Hobson_2020,CDA,CDA2,Ghilen01,Ghilen02,DG1}.
The important fact is that there does exist
a broken phase in WQG, necessary for this theory to be realistic:
the Weyl gauge boson ($\omega_\mu$) of WQG becomes massive  \cite{Ghilen01,Ghilen02} by a
Stueckelberg mechanism   and  decouples. In conclusion, non-metricity
effects are strongly suppressed by the mass of $\omega_\mu$
 and Einstein's argument does not apply. 

 In the broken phase of Weyl gauge symmetry of the WQG action, one obtains
 the  Einstein-Proca action plus
a positive cosmological constant \cite{Ghilen01,Ghilen02}. 
One can also naturally embed SM in Weyl geometry, with no new degrees of freedom (dof's) beyond SM and Weyl geometry,
while respecting all current data \cite{SMW}. One then obtains a theory of Weyl quadratic gravity
and SM, that is Weyl anomaly-free \cite{DG1}. This is possible because the
Weyl gauge symmetry is maintained in $d=4-2\epsilon$ dimensions, something allowed
only in Weyl  geometry where Weyl scalar curvature transforms Weyl gauge
covariantly.

Furthermore, there exists a generalisation of WQG in the strong-field regime,
called Weyl-Dirac-Born-Infeld (WDBI) theory of
Weyl geometry  \cite{WDBI1,WDBI2}, that includes the SM and is manifestly  Weyl gauge
invariant in {\it arbitrary $d$ dimensions} (in particular in $d=4-2\epsilon$) and
is thus  Weyl anomaly free. This theory does {\it not}
need a UV regulator (!) and it recovers {\it exactly}, in the leading order,
the WQG and SM actions already {\it regularised geometrically}  by
$\hat R^{-\epsilon}$. 
The WDBI action is special in that
both geometry (metric, $\omega_\mu$) and matter (SM) contribute on equal footing
to the new ``metric''. This removes an  intractable distinction matter-geometry
in quantum gravity in general, including string theory, of
brane (matter) placed in the $d=10$ bulk (gravity).
Beyond this matter-geometry ``unification'',
the WDBI action and its leading  approximation WQG studied in this paper,
give a unified picture, by the gauge principle, of SM internal symmetries and
(space-time) gauged dilatation symmetry.  Briefly, WQG and WDBI actions are the only
gauge theories of a space-time symmetry beyond Poincar\'e,
with {\it exact} geometric interpretation \cite{CDA}, no new parameters beyond
SM and Weyl geometry and are good candidates for a quantum gauge theory of gravity.

Are there any
experimental signatures of Weyl gauge symmetry of Weyl gauge theories of
gravity (WQG and WDBI)?
These are largely unexplored but there are notable exceptions: 1)~inflation
 where a tensor-to-scalar ratio $r$ is predicted \cite{Ghilencea:2019rqj,Ferreira:2019zzx}
 in agreement with current data;
2) the  galactic rotation curves that can have good fits in WQG \cite{Craciun:2023bmu}
with  $\omega_\mu$ as a  dark matter candidate - a geometric solution in Weyl geometry view
and a particle-like solution to this problem in the Riemannian geometry view;
and finally 3)   quantum non-locality (in particular entanglement)
may actually be a footprint of
Weyl gauge symmetry  \cite{Ghilencea:2026ugc}.

The  purpose of this paper is to explore other potential signatures
of Weyl gauge symmetry, such as  Gravitational Waves (GW). This is also motivated by recent experimental
results published by LIGO-Virgo-KAGRA (LVK) collaboration which show an increasingly favoured
trend for \textit{vector-tensor} theories beyond EH gravity \cite{LIGOScientific:2026qni,LIGOScientific:2026oim}, such as WQG and WDBI gauge theories. Since the WDBI gauge theory is a highly non-linear theory, as a first step in this direction we study the GW from its leading order, the WQG gauge  theory.

Gravitational-wave astronomy is a promising avenue for probing gravity
in the strong-field regime. GW represent ripples of the spacetime fabric
and were first predicted theoretically by Einstein \cite{1916SPAW.......688E}, while the first
indirect evidence of their existence can be traced back to the discovery of the Hulse-Taylor
binary pulsars \cite{Hulse_Taylor}. However, it was their first direct detection in a binary
black hole (BBH) merger by LIGO \cite{LIGOScientific:2016aoc} that opened a new observational window
on the universe. Recent advances made by the LVK earth-based detectors
allow a range of high energy precision tests to be carried out
\cite{LIGOScientific:2026qni,LIGOScientific:2026fcf,LIGOScientific:2026wpt,LIGOScientific:2026oim}. In particular,
the current network of detectors makes it possible to investigate the polarisation content of
GW. While EH gravity predicts only two tensor polarisation modes,
the detection of any additional vector or scalar modes would mark a strong indication
for new physics.

In this work we investigate the GW polarisations of Weyl quadratic gravity
in Weyl geometry. The paper is organised as follows. In section \ref{sec_2: Weyl
  geometry...} we  review Weyl conformal geometry in the Weyl gauge covariant
(metric!) formulation, together with its action. In
section \ref{sec_3.1: Riemannian geometry} we review geodesics in Riemannian geometry, while in
section \ref{sec_3.2: Weyl geometry} we show how to carry these arguments from Riemannian to Weyl
geometry in such formulation.
In particular we derive the geodesics deviation equation in  Weyl geometry.
In section \ref{sec_4:Polarisation modes for Einstein GR} we compute the GW
polarisations for EH action in the presence of
a positive cosmological constant $\Lambda>0$.
We develop the formalism needed and explain the
strategy that we use in de Sitter space-time. In section
\ref{sec_5: Weyl quadratic gravity} we compute the GW polarisation content for WQG on a curved background following the plan outlined in the previous section,
to find additional vector polarisation modes.
Finally, in section \ref{sec_6: Conclusions} we compare our findings against current experimental
data and draw our conclusions.

\section{Weyl geometry as a gauge theory of the Weyl group}
\label{sec_2: Weyl geometry...}

We first present a brief review of Weyl geometry and its associated action,
as a gauge theory of the Weyl group, in the Weyl gauge covariant metric formulation
\cite{DG1,CDA2} \footnote{
The formalism here can be extended to arbitrary $d$ dimensions while preserving
manifest Weyl gauge covariance/invariance, but here we work in $d = 4$ dimensions. }.

Weyl geometry is defined by classes of equivalence of the metric $g_{\mu\nu}$ and of the gauge
field of dilatations $\omega_\mu$, related by a Weyl gauge transformation:
\medskip
\begin{align}
  g_{\mu\nu} &\rightarrow g_{\mu\nu}' = \Sigma^2 g_{\mu\nu} \equiv e^{2 \lambda_D} g_{\mu\nu} \,,  &
  \omega_\mu &\rightarrow \omega_\mu' = \omega_\mu - \partial_\mu \ln \Sigma \equiv \omega_\mu -
  \partial_\mu \lambda_D \,, \label{Weyl_affine_gaugetr}
\end{align}

\medskip\noindent
where $\Sigma(x) \equiv e^{\lambda_D(x)} > 0$ is dimensionless. Here we set the Weyl
charge of $g_{\mu\nu}$ to $\tilde q = 2$. However note that for an Abelian symmetry this is a
choice, rather than a requirement. To work with an arbitrary charge $\tilde q$ for the metric,
and also restore the Weyl gauge coupling $\alpha_0$, one can simply replace $\Sigma^2
\rightarrow \Sigma^{\tilde q}$ and $\omega_\mu \rightarrow \frac{\alpha_0 \tilde q}{2} \omega_\mu$
everywhere. To complete the definition of the geometry, in addition to
\eqref{Weyl_affine_gaugetr}, one has to include the so-called "non-metricity" condition:
\medskip
\begin{equation}
  \tilde \nabla_\mu g_{\alpha\beta} + 2 \omega_\mu g_{\alpha\beta} = 0 \,, \quad \text{where} \quad
  \tilde \nabla_\mu g_{\alpha\beta} \equiv \partial_\mu g_{\alpha\beta} - \tilde
  \Gamma^\lambda_{\mu\alpha} g_{\lambda \beta} - \tilde \Gamma^\lambda_{\mu\beta} g_{\alpha\lambda} \,.
  \label{nonmetricity}
\end{equation}
\medskip\noindent
Assuming that the connection $\tilde \Gamma$ is symmetric (\ie $\tilde \Gamma^\mu_{\alpha\beta} =
\tilde \Gamma^\mu_{\beta\alpha}$), and thus torsion-free, we can solve the above equation for
$\tilde \Gamma$ by the usual techniques, finding:
\medskip
\begin{equation}
  \tilde \Gamma^\rho_{\mu\nu} = \Gamma^\rho_{\mu\nu} \Big|_{\partial_\mu \rightarrow \partial_\mu + 2
    \omega_\mu} = \Gamma^\rho_{\mu\nu} + \left( \delta^\rho_\mu \omega_\nu + \delta^\rho_\nu \omega_\mu -
  g_{\mu\nu} \omega^\rho \right) \,,
\end{equation}

\medskip\noindent
where $\Gamma$ is the usual Levi-Civita (LC) connection, $\Gamma^\rho_{\mu\nu} = \frac{1}{2}
g^{\rho\lambda} \left( \partial_\mu g_{\lambda\nu} + \partial_\nu g_{\mu\lambda} - \partial_\lambda
g_{\mu\nu} \right)$. It is straightforward to show that $\tilde \Gamma$ is invariant under
\eqref{Weyl_affine_gaugetr}. In this formulation of Weyl geometry, common in the literature
and used for almost a century, one can associate a Riemann curvature tensor $\tilde
R^\rho{}_{\sigma\mu\nu}$ to the Weyl connection $\tilde \Gamma$, which amounts to simply replacing
$\Gamma$ by $\tilde \Gamma$ in the well-known formulae of Riemannian geometry, and then
compute the Ricci tensor $\tilde R_{\mu\nu}$, and Ricci scalar $\tilde R$, respectively.
However, covariant derivatives acting on curvatures do not transform covariantly under
\eqref{Weyl_affine_gaugetr} \eg $\tilde \nabla_\mu \tilde R \rightarrow \tilde \nabla_\mu'
\tilde R' \neq \Sigma^{-2} \tilde \nabla_\mu \tilde R$. This means that the formulation itself
it is not Weyl gauge covariant, and hence not physical.

Nevertheless, a Weyl gauge covariant formulation for this geometry exists, as required for a
gauge theory. Within this framework, we define the gauge covariant derivative $\hat
\nabla_\mu$ by its action on some tensor field $T \equiv T^{\mu_1 \cdots \mu_r}{}_{\nu_1 \cdots \nu_p}$:
\medskip
\begin{equation}
  \hat \nabla_\mu T = \Big[ \tilde \nabla_\mu \big( \tilde \Gamma \big) + \tilde q_T \,
    \omega_\mu \Big]\, T \,, \label{covdev_Weyl}
\end{equation}

\medskip\noindent
where $\tilde q_T$ is the spacetime Weyl charge of $T$, \ie $T \rightarrow T' = \Sigma^{\tilde
  q_T} T$. This ensures that covariant derivatives transform  similarly to the objects on
which they act $\hat \nabla_\mu T \rightarrow \hat \nabla_\mu' T' = \Sigma^{\tilde q_T} \hat
\nabla_\mu T$. Furthermore, the Weyl gauge
covariant derivative satisfies the metricity condition, namely
$\hat \nabla_\mu g_{\alpha\beta} = 0$. Its dependence on the charge $\tilde q_T$ of the
field $T$ implies that one cannot associate a connection $\hat \Gamma$ to all  fields
as in \eqref{nonmetricity}, so although \textit{metric}, it is
\textit{non-affine}.

One can then define the Riemann tensor for Weyl geometry via the
commutator\cite{CDA2,CDA,Condeescu:2024cjh}:
\begin{equation}\label{RWG}
  \Big[ \hat \nabla_\mu, \hat \nabla_\nu \Big] V^\rho = \hat R^\rho{}_{\sigma \mu \nu} V^\sigma
  \,, 
\end{equation}

\medskip\noindent
where $V^\rho = e^\rho_a \, V^a$ is a vector with vanishing tangent space Weyl charge $q_{V^a} = 0$ (or,
equivalently, Weyl spacetime charge $\tilde q_{V^\rho} = -1$), and correspondingly the Weyl-Ricci
tensor $\hat R_{\mu\nu} = \hat R^\sigma{}_{\mu\sigma\nu}$ and Weyl-Ricci scalar $\hat R = g^{\mu\nu} \hat
R_{\mu\nu}$. One can also define the Weyl tensor $\hat C^\mu{}_{\nu\alpha\beta}$ associated to $\hat
R^\mu{}_{\nu\alpha\beta}$ and show that it is identical to its Riemannian counterpart, \ie $\hat
C^\mu{}_{\nu\alpha\beta} = C^\mu{}_{\nu\alpha\beta}$ \cite{DG1,Condeescu:2024cjh}.
The expressions  of these curvature fields
of Weyl geometry in terms of their Riemannian counterparts are shown  in Appendix \ref{App_A}. Regarding the  field strength
tensor $\hat F_{\mu\nu}$, this is  $\hat F_{\mu\nu} = \partial_\mu \omega_\nu - \partial_\nu
\omega_\mu$. Under  Weyl gauge transformations \eqref{Weyl_affine_gaugetr}, the curvature
fields behave as follows:
\begin{align}
    X &\rightarrow X \,, & X &= \hat R_{\mu\nu} \,, \hat R^\rho{}_{\sigma \mu \nu} \,, \label{tr1}\\
    X &\rightarrow \Sigma^{-2} X \,, & X &= \hat R\,, \label{tr2}\\
    X &\rightarrow \Sigma^{-4} X \,, & X &= \hat R_{\mu\nu\rho\sigma}^2 \,, \hat R_{\mu\nu}^2 \,, \hat
    C_{\mu\nu\rho\sigma}^2 \,, \hat G \,, \hat F_{\mu\nu}^2 \,, \label{tr3}
\end{align}

\medskip\noindent
where $\hat G$ is the Chern-Euler-Gauss-Bonnet term of Weyl geometry (see Appendix \ref{App_A}), which is a total
derivative in $d=4$ dimensions, but relevant in $d$ dimensions.
Also note that above the square of a tensor denotes a contraction between terms with indices
in the same position. Identities
\eqref{tr1}-\eqref{tr3} together with the properties of covariant derivatives outlined
below \eqref{covdev_Weyl} indicate that in this formulation Weyl geometry is a
covariantised version of Riemannian geometry with respect to gauged dilatations,
as expected given its symmetry.

Using the curvatures terms and their properties outlined above, we can write down the action
for a Weyl gauge theory of gravity, hereafter called \textit{Weyl quadratic gravity},
associated to Weyl geometry in $d = 4$ dimensions and invariant under
\eqref{Weyl_affine_gaugetr} \cite{SMW,Ghilen01,Ghilen02,Weyl2}:
\medskip
\begin{equation}
  S_W = \int d^4 x \sqrt{-g} \,\, \Big\{ \,\frac{1}{4! \xi^2} \hat R^2 - \frac{1}{\eta^2} \hat
  C_{\mu\nu\rho\sigma} \hat C^{\mu\nu\rho\sigma} - \frac{1}{4 \alpha_0^2} \hat F_{\mu\nu} \hat F^{\mu\nu} +
  \hat G \Big\} \,. \label{full_Weyl_action}
\end{equation}

\medskip\noindent
Here $\xi, \eta, \alpha_0 < 1$ are dimensionless (perturbative) couplings. This action
has a Weyl gauge current $j_\mu \propto \hat \nabla_\mu \hat
R$ which is covariantly conserved (\ie $\hat \nabla^\mu j_\mu = 0$), and extends the analogous
current found in globally scale-invariant theories \cite{F1,F2,F3,F4,Bellido}. Further,
action \eqref{full_Weyl_action} undergoes a Stueckelberg breaking of Weyl gauge symmetry,
whereby $\omega_\mu$ acquires a mass $m_\omega$, after ``eating'' the would-be-Goldstone
of gauged dilatations, $\ln\phi$. This is  the same scalar field that linearises
the $\hat R^2$ term in the action, by replacing $\hat R^2 \rightarrow - 2 \phi^2 \hat R - \phi^4$ in \eqref{full_Weyl_action}, where $\phi$ is the scalar mode propagated by $\hat R^2$; then
setting $\phi^2 \rightarrow \langle \phi^2 \rangle$  generates all  mass scales of the theory \cite{SMW,review}:
\begin{equation}
    \Lambda \equiv \frac{\langle \phi^2 \rangle}{4}\,, \qquad \quad  M_p^2 \equiv \frac{2 \Lambda}{3 \xi^2} \,, \qquad \quad m_\omega^2 \equiv 6 \alpha_0^2 M_p^2 \,, \label{mass_scales} 
\end{equation}
where $\Lambda$ is the cosmological constant and $M_p$ the Planck mass.

Note that $m_\omega$ depends on the Weyl gauge coupling $\alpha_0$, which is a parameter of the theory. One possibility is $\alpha_0 \sim \mathcal{O}(1)$, which means that the Weyl gauge field $\omega_\mu$ acquires a mass $m_\omega \sim M_p$. In this case, at low energies, $\omega_\mu$ decouples, and the theory reduces to
Riemannian geometry with the Einstein–Hilbert action and a cosmological constant $\Lambda >
0$ \cite{Ghilen01,Ghilen02}. Nonetheless, the Weyl gauge coupling $\alpha_0$ is largely unconstrained\footnote{Of the SM spectrum, only the Higgs field can couple to $\omega_\mu$
(through its Weyl gauge invariant kinetic term), with coupling $\alpha_0\sim m_\omega/M_p$
which is negligible for the scenario discussed here of ultralight $m_\omega$. In principle it is also
possible to have a gauge kinetic mixing of $\omega_\mu$ to the hypercharge gauge field of SM,
in which case the mass of $Z$ boson receives corrections from  massive $\omega_\mu$. Precision measurements of the $Z$ boson mass apparently place a lower bound on  $\alpha_0$, \ie $\alpha_0 \geq 2.17 \times 10^{-15} \sin \chi$, where $\sin \chi$ is the
gauge-kinetic mixing angle \cite{SMW}. However $\omega_\mu$ is C-even while the photon is C-odd, so the gauge kinetic mixing coupling can be naturally tuned to zero  \cite{tHooft:1979rat} 
on the grounds of C violation in the gauge sector \cite{Hayashi:1978xu,Hayashi:1976uz}
and $\alpha_0$ can be even smaller.},
hence it can have ultra-weak values, $\alpha_0 \ll 1$, and thus $m_\omega \ll M_p$.
In this regime Weyl quadratic gravity remains a vector-tensor  gauge
theory  to low energies, with the effects induced by $\omega_\mu$ largely evading experimental detection due to ultraweak coupling.
One notable exception to this are atomic clocks. Current earth-based experiments can be used to place a lower bound on the mass of ultra-light Weyl gauge boson\footnote{The lower bound on the mass of the Weyl gauge boson, $m_\omega \sim$ few TeV, found in \cite{Latorre:2017uve} was
derived from a Bhabha scattering process mediated by $\omega_\mu$ via a coupling
that does not exist in Weyl conformal geometry \cite{SMW}.}
$m_\omega \gtrsim 10^{-14}$ eV \cite{Khodadi:2026zoi}, which in turn places a lower bound on the Weyl gauge coupling $\alpha_0 \gtrsim 10^{-42}$. These bounds are also compatible with constraints from stellar-mass black holes superradiance \cite{Cardoso:2018tly}. Another notable exception, which allows us to detect $\omega_\mu$ in principle and thus constrain $\alpha_0$, are GW,
of which the Weyl gauge boson would provide the vector polarisation modes. This is discussed shortly.

A non-linear generalisation of this action is the WDBI gauge theory \cite{WDBI1,WDBI2} of Weyl geometry, that is valid in strong-field regime, with an action $S_{WDBI}=\sqrt{\det G_{\mu\nu}}$, where
$G_{\mu\nu}= a_0 \hat R g_{\mu\nu}+a_1 \hat R_{\mu\nu}+a_2\hat F_{\mu\nu}$ is the space-time metric 
(ignoring here the SM contributions to the metric). $S_{WDBI}$
can be expanded in a power series in $1/a_0\sim \xi^2\ll 1$, the first order of which
is exactly WQG action \eqref{full_Weyl_action}. Here we study its GW polarisation modes.

\section{Geodesics and geodesic deviation equation}
\label{sec_3: Geodesics}

\subsection{Riemannian geometry}
\label{sec_3.1: Riemannian geometry}

In order to discuss GW in Weyl geometry, let us first review the case of geodesics in the Riemannian
case. An affinely-parametrised geodesic is defined to be a curve whose tangent
vector is parallel transported along itself \cite{Wald:1984rg}, \ie
\medskip
\begin{equation}
    X^\nu \nabla_\nu X^\mu = 0 \,,
\end{equation}

\medskip\noindent
which in component form reads:
\medskip
\begin{equation}
  \frac{\partial X^\mu}{\partial \tau} + \Gamma^\mu_{\nu\rho} X^\nu X^\rho = 0 \quad
  \Leftrightarrow \quad \frac{\partial^2 x^\mu}{\partial \tau^2} + \Gamma^\mu_{\nu\rho}
  \frac{\partial x^\nu}{\partial \tau} \frac{\partial x^\rho}{\partial \tau}= 0\,.
  \label{geodesiceq}
\end{equation}

\medskip\noindent
To see how two neighbouring geodesics behave with respect to one another, we consider a
one-parameter family of geodesics $\gamma_s(\tau)$. This family amounts to a hypersurface
within the manifold which can be described by a set of coordinates $x^\mu(\tau;s)$, where
$\tau$ is the affine parameter along each geodesic, while the remaining parameter $s$ labels
the different geodesics within the hypersurface. The tangent vector along the geodesic is
given by:
\medskip
\begin{equation}
    X^\mu = \frac{\partial x^\mu}{\partial \tau} \bigg|_s \,, \label{tangentvector}
\end{equation}

\medskip\noindent
while the deviation vector connecting two neighbouring geodesics is:
\medskip
\begin{equation}
    S^\mu = \frac{\partial x^\mu}{\partial s} \bigg|_\tau \,. \label{deviationvector}
\end{equation}

\medskip\noindent
In this setup, the geodesic deviation equation, \ie the tidal acceleration between two
neighbouring geodesics, takes the following form:
\medskip
\begin{equation}
  \frac{d^2 S^\mu}{d \tau^2} = R^\mu{}_{\nu \rho \sigma} X^\nu X^\rho S^\sigma \,,
  \label{geodesicdeveq}
\end{equation}

\medskip\noindent
where $d / d \tau = \left( \partial x^\mu / \partial \tau \right) \nabla_\mu = X^\mu \nabla_\mu$.

GW detectors are designed to detect the relative displacement of two space-like separated
freely falling test masses. For an earth-based detector, we can choose to work in the
\textit{local detector frame} (see Figure \ref{figure_geod_dev} below). This means that,
prior to the arrival of any GW, $X^\mu = (1,0,0,0)$. Gravitational waves occur as fluctuations
of the metric, \ie $g_{\mu\nu} = \eta_{\mu\nu} + h_{\mu\nu}$, then $\Gamma[\eta_{\mu\nu}] \rightarrow \Gamma[\eta_{\mu\nu} + h_{\mu\nu}]$, which means that geometry is locally modified from the flat space-time approximation of the detector.
This implies that the tangent vector now
becomes $X^\mu = (1,0,0,0) + O(h)$. Writing the deviation vector as $S^\mu = S_0^\mu + S_1^\mu + \cdots$, where $S_n^\mu = O(h^n)$, eq.\eqref{geodesicdeveq} at leading order in fluctuations reads:
\begin{equation}
    \frac{d^2 S_1^i}{d t^2} = \left[ R^i{}_{00j} \right]^{(1)} S_0^j \,. \label{geodesic1}
\end{equation}
Note that $\left[ R^i{}_{00j} \right]^{(1)}$ is the $O(h)$ part of the corresponding Riemann tensor, $S_0^i$ represents the original separation of the test masses, while $S_1^j$ the extra displacement induced by the GW passing.

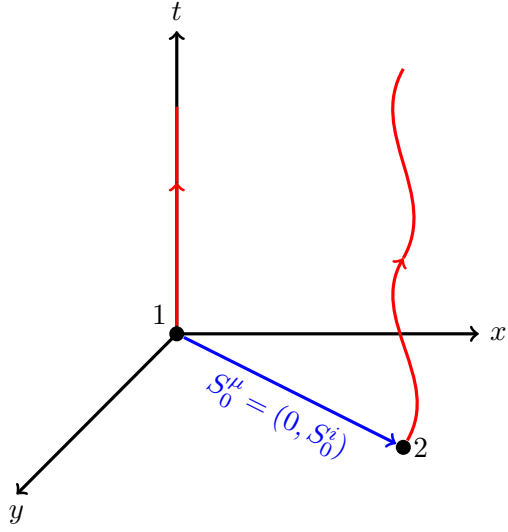
\begin{figure}[h!] 
    \centering
    \scalebox{1}{
      \begin{tikzpicture}
        \draw[->, very thick] (0,0)--(4,0) node[right]{$x$};
        \draw[->, very thick] (0,0)--(0,4) node[above]{$t$};
        \draw[->, very thick, rotate=135] (0,0)--(0,3) node[below]{$y$};
        \node [circle, fill, inner sep=2pt] at (0,0) (a) {};
        \node[above left] at (0,0) {$1$};
        \node [circle, fill, inner sep=2pt] at (3,-1.5) (b) {};
        \node [right] at (3,-1.5) {$2$};
        \draw[->, very thick, blue] (a)--(b) node[midway,sloped,below]{$S_0^\mu=(0,S_0^i)$};
        \draw[very thick, red, ->] (a)--(0,2);
        \draw[very thick, red] (0,2)--(0,3);
        \draw [very thick, red, ->] (b) to[out=+60,in=-120] (3,1); 
        \draw [very thick, red] (3,1) to[out=60,in=-120] (3,3.5);
    \end{tikzpicture}}
    \caption{The relative (space-like) displacement of two neighbouring geodesics in the initial rest frame of the first test mass (\ie \textit{local detector frame}). }
    \label{figure_geod_dev}
\end{figure}

\subsection{Weyl geometry}
\label{sec_3.2: Weyl geometry}

Since Weyl geometry is a covariantised version of
Riemannian geometry with respect to gauged dilatations,
the equation of a geodesic in Weyl geometry is simply a covariantised version of the
geodesic in Riemannian case and its equation 
is given by
($e_a^\mu$ has charge $\tilde q=-1$):
\medskip
\begin{equation}
  X^\mu \hat \nabla_\mu X^\nu = 0 \quad \Leftrightarrow \quad X^\mu \Big[ \tilde \nabla_\mu -
    \omega_\mu \Big] X^\nu = 0 \,. \label{Weyl_geodesic}
\end{equation}

\medskip\noindent
Next, since $X^\mu$ and $S^\mu$ are coordinate vector fields, they
commute\footnote{We can always shift the affine parameter $\tau \rightarrow \tau' = A(s)
  \tau + B(s)$ such that the new coordinate vector fields ($X^\mu \rightarrow X'^\mu$ and
  $S^\mu \rightarrow S'^\mu$) obey the  relations: $g^{\mu\nu} X'_\mu X'_\nu = \pm 1$,
  provided $A(s) = \left( X \cdot X \right)^{-\frac{1}{2}}$, and $g^{\mu\nu} X'_\mu S'_\nu = 0$ if we
  choose $B'(s) = - \frac{S \cdot X}{X \cdot X}$. Note that the normalisation condition for
  $X_\mu$ is required to ensure that $X^\mu S_\mu$ vanishes at any point along the geodesic, and
  thus the two vectors remain orthogonal to one another.}:
\medskip
\begin{equation}
  \left[ X, S \right] = 0 \quad \Leftrightarrow \quad X^\mu \nabla_\mu S^\nu = S^\mu \nabla_\mu
  X^\nu \,.
\end{equation}

\medskip\noindent
This equation is true in Riemannian geometry and more generally for any metric-compatible
connection, which means it is also true in Weyl geometry in its metric formulation
\medskip
\begin{equation}
    X^\mu \hat \nabla_\mu S^\lambda = S^\mu \hat \nabla_\mu X^\lambda \,. \label{commutatorWeyl}
\end{equation}

\medskip\noindent
We also checked in Appendix~\ref{App_B} that this holds true.
Further, if we apply $\hat
\nabla_X \equiv X^\nu \hat \nabla_\nu$ to both sides of the above equation, then:
\medskip
\begin{equation}
  X^\nu \hat \nabla_\nu \big( X^\mu \hat \nabla_\mu S^\lambda \big) = X^\nu \hat \nabla_\nu
  \big( S^\mu \hat \nabla_\mu X^\lambda \big) \,.
\end{equation}

\medskip\noindent
With $d/{d \tau} = X^\mu \hat \nabla_\mu$, the last equation can be recast as follows:
\medskip
\begin{align}
  \frac{d^2 S^\lambda}{d \tau^2} &= X^\nu \hat \nabla_\nu \big( S^\mu \hat \nabla_\mu X^\lambda
  \big) \nonumber \\
  &= X^\nu \big( \hat \nabla_\nu S^\mu \big) \big( \hat \nabla_\mu X^\lambda \big) + X^\nu
  S^\mu \big( \hat \nabla_\nu \hat \nabla_\mu X^\lambda \big) \nonumber \\
  &= S^\nu \big( \hat \nabla_\nu X^\mu \big) \big( \hat \nabla_\mu X^\lambda \big) + X^\nu
  S^\mu \big( \big[ \hat \nabla_\nu , \hat \nabla_\mu \big] + \hat \nabla_\mu \hat \nabla_\nu
  \big) X^\lambda \\
  &= S^\nu \hat \nabla_\nu \big( X^\mu \hat \nabla_\mu X^\lambda \big) + X^\nu S^\mu \hat
  R^\lambda{}_{\rho\nu\mu} X^\rho \nonumber \\ 
    &= \hat R^\lambda{}_{\rho\nu\mu} X^\rho X^\nu S^\mu \,, \nonumber
\end{align}

\medskip\noindent
where we used definition (\ref{RWG}) in the last line. In going from the first to the second line we used the Leibniz property of the covariant
derivative, from the second to the third line we used \eqref{commutatorWeyl}, and in the fourth
line we collected inside the bracket the first and the third term from the third line to recover
the geodesic equation \eqref{Weyl_geodesic} for $X^\lambda$. Following the same reasoning as in
the previous section, the geodesic deviation equation at leading order in fluctuations is
given by (with (\ref{a1})):
\medskip
\begin{equation}
  \frac{d^2 S_1^i}{d t^2} = \big[ \hat R^i{}_{00j} \big]^{(1)} S_0^j = \big( \big[
    R^i{}_{00j} \big]^{(1)} + \delta^i_j \, \partial_0 \omega_0 + \partial_j \omega^i \big)
 \,\, S_0^j \,. \label{Weylgeodesic1}
\end{equation}

\medskip
This equation is a generalisation of the Riemannian case (\ref{geodesic1}) which  now
receives corrections due to the  Weyl gauge field of dilatations $\omega_\mu$.
If $\omega_\mu=0$, we recover (\ref{geodesic1}).

\section{Polarisation modes of GW for Einstein-Hilbert dS}
\label{sec_4:Polarisation modes for Einstein GR}

For our later study of GW in Weyl geometry, in this section we review the GW polarisation
modes for Einstein-Hilbert gravity in de Sitter (dS) background.
We consider here the Einstein-Hilbert action with a cosmological constant $\Lambda > 0$:
\begin{equation}
  S = - \frac{M_p^2}{2} \int d^4 x \sqrt{-g} \, \big( R + 2 \Lambda \big) \,.
  \label{Einstein_dS_action}
\end{equation}
The  equations of motion are:
\begin{equation}
    G_{\mu\nu} = R_{\mu\nu} - \frac{1}{2} g_{\mu\nu} R = \Lambda g_{\mu\nu} \,. \label{Einstein_dS_EOM}
\end{equation}
We are interested in studying linear order (quantum) perturbations around some background
which satisfies \eqref{Einstein_dS_EOM} above. Classically, one such background is a dS universe, whose line element can be written in planar coordinates as follows:
\begin{equation}
    d s^2 = d t^2 - a^2(t) \delta_{ij} d x^i d x^j \,, \label{dS_1}
\end{equation}
where $a(t) = e^{H_0 t}$ and $H_0^2 = \Lambda/3 > 0$. It is more convenient and useful
for the subsequent computation to replace the cosmic time $t$ with conformal time $\eta$ using
$dt = a(\eta) d \eta$, the line element in \eqref{dS_1} now reading:
\begin{align}
  d s^2 &= a^2(\eta) \big( d \eta^2 - \delta_{ij} d x^i d x^j \big) \,,  \qquad \text{with
  }  a(\eta) = - \frac{1}{H_0 \eta} \,.
\end{align}

\subsection{Gauge symmetry}
At linear order, the theory inherits a gauge symmetry from the full (non-linear) theory. To
see how this unfolds, and accounting for the background, we first decompose the full metric as
follows:
\begin{equation}
    g_{\mu\nu} = a^2(\eta) \big( \eta_{\mu\nu} + h_{\mu\nu} \big) \,,
\end{equation}

\medskip\noindent
with notation $\bar g_{\mu\nu} = a^2 \eta_{\mu\nu}$ and $\delta g_{\mu\nu} = a^2 h_{\mu\nu}$. Under a change
of coordinates $x^\mu \rightarrow x^\mu + \xi^\mu$ the full metric $g_{\mu\nu}$ transforms as:
\medskip
\begin{equation}\label{tra}
  g_{\mu\nu} \rightarrow g_{\mu\nu} - \nablabar_\mu \xi_\nu - \nablabar_\nu \xi_\mu = g_{\mu\nu} -
  \xi^\rho \partial_\rho \bar g_{\mu\nu} - \bar g_{\rho\nu} \partial_\mu \xi^\rho - \bar g_{\mu\rho}
  \partial_\nu \xi^\rho \,,
\end{equation}

\medskip\noindent
with $\overline\nabla \equiv \overline\nabla \left[ \bar g_{\mu\nu} \right]$. At linear order in fluctuations (\ref{tra}) becomes:
\smallskip
\begin{equation}
  a^2 h_{\mu\nu} \rightarrow a^2 h_{\mu\nu} - \xi^\rho \partial_\rho \bar g_{\mu\nu} - \bar
  g_{\rho\nu} \partial_\mu \xi^\rho - \bar g_{\mu\rho} \partial_\nu \xi^\rho \,.
  \label{gaugesymmetry_EinsteindS}
\end{equation}

\smallskip\noindent
To address this redundancy one can either choose an appropriate gauge to work in, or,
alternatively, one can recast the equations of motion in terms of gauge invariant variables
\cite{Bardeen,Flanagan_2005}. Performing a gauge invariant decomposition (in Scalar, Vector, and Tensor modes, or combinations thereof) is the only way to show whether a mode is a real (observable) physical wave carrying positive energy (and not a ghost). This is the approach that we follow here.

\subsection{Scalar-Vector-Tensor decomposition}
\label{sec_4.2:SVT - Einstein dS}

To do this, we first split $h_{\mu\nu}$ into its irreducible components under spatial rotations:
\begin{align}
     h_{00} &= 2 \psi \,, \\
     h_{0i} &= \beta_i + \partial_i \gamma \,, \\
     h_{ij} &= -2 \, \theta \, \delta_{ij} + \Big( \partial_i \partial_j -\frac{1}{3}
     \delta_{ij} \partial_m \partial_m \Big)\lambda + \frac{1}{2} \Big( \partial_i \epsilon_j
     + \partial_j \epsilon_i \Big) + h_{ij}^{TT} \,, \label{hij}
 \end{align}

\medskip\noindent
where: $ \partial_i \beta^i = 0 \,, \partial_i \epsilon^i = 0 \,,\partial^j h^{TT}_{ij} = 0 \,,
\delta^{ij} h_{ij}^{TT} = 0$. Note that $h^{\alpha\beta} = \eta^{\alpha\mu} \eta^{\beta\nu} h_{\mu\nu}$,
thus $h^{00} = h_{00}$, $h^{0i} = - h_{0i}$ and $h^{ij} = h_{ij}$. To ensure the
uniqueness of the SVT decomposition \cite{Jaccard_2013}, we require $\gamma \rightarrow 0 \,,
\epsilon_i \rightarrow 0 \,, \lambda \rightarrow 0 \,, \nabla^2 \lambda \rightarrow 0$ as $r
\rightarrow \infty$. Additionally, we require that $h_{\mu\nu} \rightarrow 0$ as $r \rightarrow
\infty$. The 10 original dof's of $h_{\mu\nu}$ are now split into 4 scalar
dof's $\big\{ \psi, \gamma, \lambda, \theta \big\}$, 4 vector dof's
$\big\{ \beta_i, \epsilon_i \big\}$, and 2 tensor dof's in the
transverse-traceless (TT) spatial $h_{ij}^{TT}$. However, when subjected to
\eqref{gaugesymmetry_EinsteindS}, we are left with $10-4=6$ possible polarisation modes (\ie
independent dof's).  

For $\xi^\mu$ we proceed along the same lines writing it as follows:
\smallskip
\begin{equation}
  \xi^\mu = \big( A, B^i + \partial^i C \big) \,, \qquad \text{with } \partial_i B^i = 0
  \,.
\end{equation}

\smallskip\noindent
As before, we raise/lower indices using the background metric: $\xi_\mu = \bar g_{\mu\nu} \xi^\nu
\equiv a^2 \big( A, B_i + \partial_i C \big)$, and $B_i = - B^i$, $\partial_i C = - \partial^i C$. The gauge transformations \eqref{gaugesymmetry_EinsteindS} can be
written for each individual dof as follows:
\begin{equation}
    \begin{aligned}
      \psi &\rightarrow \psi - \hh A - A' \,,  &  \qquad \qquad\beta_i &\rightarrow \beta_i
      + B'_i \,,  &  \qquad \qquad h_{ij}^{TT} &\rightarrow h_{ij}^{TT} \,, \\
      \theta &\rightarrow \theta - \hh A - \frac{1}{3} \partial_m \partial_m C \,,  &
      \epsilon_i &\rightarrow \epsilon_i + 2 B_i \,, \\
        \lambda &\rightarrow \lambda + 2 C \,, \\
        \gamma &\rightarrow \gamma + C' - A \,,
    \end{aligned} \label{gauge_inv_dS}
\end{equation}
where a prime  stands for a derivative with respect to conformal time $\eta$, and $\hh = a' / a
= - 1 / \eta$. This in turn enables us to construct gauge invariant quantities by combining the
elementary fields:
\begin{equation}
    \begin{aligned}
      S_1 &= - \theta - \frac{1}{6} \partial_m \partial_m \lambda + \hh\,\, \Big( \gamma -
      \frac{1}{2} \lambda' \Big) \,,  & \qquad V_i &= \beta_i - \frac{1}{2}
      \epsilon_i' \,,  & \qquad h_{ij}^{TT} &= h_{ij}^{TT} \,, \\
      S_2 &= - \psi + \gamma' - \frac{1}{2} \lambda'' + \hh \,\,\Big( \gamma - \frac{1}{2}
      \lambda' \Big) \,.
    \end{aligned}
\end{equation}
This means that the six polarisation modes are now split as two scalar modes, two vector modes
(since $\partial_i V^i = 0$) and two tensor modes. Note that in the limit $a \rightarrow 1$ we
recover the corresponding flat space invariants since $\hh \rightarrow 0$. This allows us to
recast the Riemann tensor of  the geodesic deviation equation
\eqref{geodesic1} in terms of invariants, as follows:
\begin{equation}\label{EHdS}
  \big[ R^{i}{}_{00j} \big]^{(1)} = \partial_i \partial _j S_2 - \ddot S_1 \delta_{ij} +
  \frac{1}{2}\, \big( \partial_i \dot V_j + \partial_j \dot V_i \big) - \frac{1}{2} \ddot
  h_{ij}^{TT} \,,
\end{equation}
where a dot stands for a derivative with respect to coordinate time $t$.

\subsection{Polarisation modes}

Having already written everything in terms of gauge invariant
variables, for  further simplifications,  we can choose to do our computations
in a particular gauge in which expressions
\eqref{gauge_inv_dS} for each gauge invariant quantity become simpler.
The final result, in terms  of gauge invariant quantities, must hold not only in the chosen gauge, but
in any gauge \cite{Maggiore:2018sht}. To avoid any spurious gauge modes, we choose a gauge
which removes gauge freedom completely. One simple choice to achieve this is:
\begin{equation}
    \lambda = \gamma = \epsilon_i = 0 \label{GF1}\,,
\end{equation}
such that $\lambda = 0$ fixes $C$, $\gamma = 0$ fixes $A$, (with $C$  already fixed), while
$\epsilon_i = 0$ fixes $B_i$.

The equations of motion of (\ref{Einstein_dS_action})
are trivially satisfied at the background level, while at linear order
in fluctuations they are given by:
\medskip
\begin{equation}
    R_{\mu\nu}^{(1)} = - 3 \hh^2 h_{\mu\nu} \,, \qquad R^{(1)} = 0 \,.
\end{equation}

\medskip\noindent
The (00), (0i), (ij) components of the former, and the latter equation, respectively, read:
\medskip
\begin{align}
  0 &= - \partial_m \partial_m S_2 + 3 S_1'' - 3 \hh S_2' + 3 \hh S_1' - 6 \hh^2 S_2 \,,
  \label{00}\\
  0 &= 2 \partial_i S_1' + \frac{1}{2} \partial_m \partial_m V_i - 2 \hh \partial_i S_2 + 2
  \hh V_i' + 10 \hh^2 V_i \,, \label{0i} \\
  0 &= \delta_{ij} \left[ - S_1'' + \partial_m \partial_m S_1 + 6 \hh S_2 + S_1 \left( 6 \hh^2
    - 6 \hh \right) - 5 \hh S_1' + \hh S_2' \right] + \partial_i \partial_j (S_1 + S_2)
  \nonumber \\
  &- \frac{1}{2} \left( \partial_0 \partial_0 - \partial_m \partial_m \right) h_{ij}^{TT} +
  \frac{1}{2} \left( \partial_i V_j' + \partial_j V_i' \right) + \hh \left( \partial_i V_j +
  \partial_j V_i \right) - \hh \partial_0 h_{ij}^{TT} \,, \label{ij} \\
  0 &= - \partial_m \partial_m S_2 - 2 \partial_m \partial_m S_1 + 3 S_1'' + 9 \hh S_1' - 3
  \hh S_2' + 9 \hh (S_1 - S_2) - 3 \hh^2 ( 3 S_1 + S_2) \,, \label{R_eq}
\end{align}
\medskip\noindent
where we also introduce $\partial_0 \equiv \partial/\partial \eta$. If we now contract $\partial_i$ into \eqref{0i}, we obtain:
\medskip
\begin{equation}
  \partial_i \partial_i \left( S_1' - \hh S_2 \right) = 0 \Rightarrow S_1' = \hh S_2 \,.
  \label{eq1}
\end{equation}

\medskip\noindent
Using this into \eqref{00} then
\begin{equation}
    \partial_m \partial_m S_2 = 0 \Rightarrow S_2 = 0\,, \label{eq2}
\end{equation}

\medskip\noindent
which in turn makes $S_1' = 0$. These results are used to recast \eqref{R_eq} as follows:  
\medskip
\begin{equation}
    \partial_m \partial_m S_1 + \frac{9}{2 \eta^2} \left( \eta + 1 \right) S_1 = 0 \,,
\end{equation}

\medskip\noindent
which, under the boundary conditions outlined below eq.\eqref{hij}, renders $S_1 =
0$.

With the scalar sector now solved, we use a similar strategy for the remaining two
invariants. Contracting by $\partial_i$ in \eqref{ij}, we find:
\medskip
\begin{equation}
    V_i' + 2 \hh V_i = 0 \,.
\end{equation}

\medskip\noindent
Using  this back into \eqref{0i}, we obtain:
\medskip
\begin{equation}
    \partial_m \partial_m V_i + 12 \hh^2 V_i = 0 \,,
\end{equation}

\medskip\noindent
where boundary conditions again force $V_i = 0$. This means that \eqref{ij} now becomes:
\medskip
\begin{equation}
  \left( \partial_0 \partial_0 - \partial_m \partial_m + 2 \hh \partial_0 \right) h_{ij}^{TT} =
  0 \,. \label{wave_dS}
\end{equation}

\medskip\noindent
In the flat space-time limit (\ie $\hh \rightarrow 0$), this equation
reduces to the standard plane wave equation for the two tensor polarisation modes
of $h_{ij}^{TT}$. This equation will be extended  to the case of Weyl quadratic gravity.

\section{Polarisation modes of GW in Weyl quadratic gravity}
\label{sec_5: Weyl quadratic gravity}

To find the GW polarisation modes  in Weyl quadratic gravity,
we use the approach of the previous section. We 
consider here a simplified version of WQG action in \eqref{full_Weyl_action}\footnote{The topological term $\hat G$ is a total
derivative in $d = 4$ and thus it does not contribute to the equations of motion. The other term, $\hat C_{\mu\nu\rho\sigma}^2$, being identical to its Riemannian counterpart, can be shown that, with appropriate boundary
conditions imposed on the metric, it is equivalent to the Einstein-dS
case \cite{Maldacena:2011mk,Hell:2023rbf}. For this reason, we expect the inclusion of this term to not alter the structure of the equations of motion for fluctuations in dS background, see later eqs. \eqref{TV} and \eqref{TT}.}:
\medskip
\begin{equation}
  S = \int d^4x \sqrt{-g}\,\, \Big( d_1 \hat R^2 + d_2 \hat F_{\mu\nu} \hat F^{\mu\nu} \Big) \,.
  \label{S} 
\end{equation}

\medskip\noindent
To simplify notation, we introduced $d_1$ and $d_2$ which stand for the dimensionless coefficients in action \eqref{full_Weyl_action}, namely $d_1 \equiv 1/(4!\,\xi^2)$, $d_2 \equiv -1/(4\alpha_0^2)$.
The equations of motion take the following form
\cite{CDA2}:
\medskip
\begin{align}
  d_1 \Big( 2 g_{\mu\nu} \hat \Box \hat R - 2 \hat \nabla_\nu \hat \nabla_\mu \hat R + 2 \hat
  R_{\mu\nu} \hat R - \frac{1}{2} g_{\mu\nu} \hat R^2 \Big) + d_2 \Big( 2 \hat F_{\mu\rho} \hat
  F_\nu{}^\rho - \frac{1}{2} g_{\mu\nu} \hat F_{\rho\sigma} \hat F^{\rho\sigma} \Big) &= 0 \,,
  \label{WEOM} \\
    3 d_1 \hat \nabla_{\mu} \hat R - d_2 \hat \nabla_\nu \hat F^\nu{}_\mu &= 0 \,. \label{BEOM}
\end{align}

\medskip\noindent
We proceed as in previous section and linearise these equations about some background:
\medskip
\begin{align}
    g_{\mu\nu} &= \bar g_{\mu\nu}+ \delta g_{\mu\nu} \,, \label{gexpansion} \\
    \omega_\mu &= 0 +\delta \omega_\mu \,, \label{omegaexpansion}
\end{align}

\medskip\noindent
where $\bar g_{\mu\nu}$ is a general (curved) background metric; we also consider
$\omega_\mu$ to be a  small quantum fluctuation with no background part ($\bar\omega_\mu=0$ to
avoid Lorentz symmetry violation).

Thus, at the background level the equations of motion read:
\medskip
\begin{align}
  2 \bar g_{\mu\nu} \overline \Box \bar R - 2 \nablabar_\nu \nablabar_\mu \bar R + 2 \bar
  R_{\mu\nu} \bar R - \frac{1}{2} \bar g_{\mu\nu} \bar R^2 &= 0 \,, \label{background1} \\
    \partial_\mu \bar R &= 0 \label{background2} \,.
\end{align}

\medskip\noindent
Note that background curvatures (\ie $\bar R$, $\bar R_{\mu\nu}$) do not depend on $\omega_\mu$ since $\bar \omega_\mu = 0$. The last equation implies that $\bar R$ is a spacetime constant, while \eqref{background1}
reduces to:
\medskip
\begin{equation}
    \bar R \Big( \bar R_{\mu\nu} - \frac{1}{4} \bar g_{\mu\nu} \bar R \Big) = 0 \,. 
\end{equation}

\medskip\noindent
This means that maximally symmetric spacetimes, namely $\bar R_{\mu\nu} = - \Lambda \bar
g_{\mu\nu} \Leftrightarrow \bar R = - 4 \Lambda$, are the only valid backgrounds
to consider. Although a flat background (\ie $\bar R = 0$) is a valid possibility, it raises questions regarding both stability and phenomenological viability \cite{Fan:2026lms}.
In this work we consider a dS background with $\Lambda > 0$.

\subsection{Gauge symmetry}
The full theory exhibits three kinds of invariance:
\begin{itemize}
    \item Diffeomorphism invariance $x^\mu \rightarrow x^\mu + \xi^\mu$:
    \begin{align}
      \delta_\xi g_{\mu\nu} &= - \nablabar_\mu \xi_\nu - \nablabar_\nu \xi_\mu = - \xi^\rho
      \partial_\rho \bar g_{\mu\nu} - \bar g_{\rho\nu} \partial_\mu \xi^\rho - \bar g_{\mu\rho}
      \partial_\nu \xi^\rho \,, \\
      \delta_\xi \omega_\mu &= - \xi^\nu \partial_\nu \omega_\mu - \omega_\nu \partial_\mu \xi^\nu
      \,.
    \end{align}
    
    \item Gauged dilatation invariance:
    \begin{align}
        \delta_\lambda g_{\mu\nu} &= 2 \lambda_D g_{\mu\nu} \,, \\
        \delta_\lambda \omega_\mu &= - \partial_\mu \lambda_D \,.
    \end{align} 
    
  \item Gauged covariant diffeomorphism invariance (\ie diffeomorphism generated by $\xi$ +
    gauged dilatation with $\lambda_D = -\omega_\rho \xi^\rho$) \cite{Condeescu:2024cjh,CDA2}:
    \begin{align}
        \hat\delta_\xi g_{\mu\nu} &= - \hat\nabla_\mu \xi_\nu - \hat\nabla_\nu \xi_\mu \,, \\
        \hat\delta_\xi \omega_\mu &= F_{\mu\nu} \xi^\nu \,.
    \end{align}
\end{itemize}
This means that the combined gauge invariance follows from:
\begin{align}
  \left( \delta g_{\mu\nu} \right) &\rightarrow \left( \delta g_{\mu\nu} \right) - \xi^\rho
  \partial_\rho \bar g_{\mu\nu} - \bar g_{\rho\nu} \partial_\mu \xi^\rho - \bar g_{\mu\rho} \partial_\nu
  \xi^\rho + 2 \lambda_D \bar g_{\mu\nu} \,, \label{gaugetransfh1}\\
    \omega_\mu &\rightarrow \omega_\mu - \partial_\mu \lambda_D \label{gaugetransfomega1}\,.
\end{align}

\subsection{Scalar-Vector-Tensor decomposition}
\label{sec_5.2: SVT decomposition}

We use the same conventions as in section~\ref{sec_4.2:SVT - Einstein dS}
and follow a similar strategy. For the additional $\omega_\mu$ term, we
use the following decomposition:
\medskip
\begin{equation}
  \omega_\mu = \left( \alpha, \rho_i + \partial_i \sigma \right) \quad \Rightarrow \quad
  \omega^\mu = \bar g^{\mu\nu} \omega_\nu \equiv a^{-2} \left( \alpha, \rho^i + \partial^i \sigma
  \right) \,,
\end{equation}

\medskip\noindent
with $\partial_i \rho^i = 0$, $\rho^i = - \rho_i$ and $\partial^i \sigma = -\partial_i
\sigma$. The gauge transformations \eqref{gaugetransfh1} and \eqref{gaugetransfomega1} for the
elementary fields are given below (with additional effect from
gauged dilatations $\lambda_D$; compare to \eqref{gauge_inv_dS}):
\begin{equation}
    \begin{aligned}
      \psi &\rightarrow \psi - \hh A - A' + \lambda_D \,,  & \qquad \qquad \beta_i
      &\rightarrow \beta_i + B'_i \,,  & \qquad \qquad h_{ij}^{TT} &\rightarrow h_{ij}^{TT} \,, \\
      \theta &\rightarrow \theta - \hh A - \frac{1}{3} \partial_m \partial_m C + \lambda_D
      \,,  &  \epsilon_i &\rightarrow \epsilon_i + 2 B_i \,, \\
        \lambda &\rightarrow \lambda + 2 C \,,  &  \rho_i &\rightarrow \rho_i \,, \\
        \gamma &\rightarrow \gamma + C' - A \,, \\
        \alpha &\rightarrow \alpha - \lambda_D' \,, \\
        \sigma &\rightarrow \sigma - \lambda_D \,.
    \end{aligned} 
\end{equation}
This allows us to define a new set of gauge invariant variables:
\begin{equation}
    \begin{aligned}
      S_1 &= - \theta - \frac{1}{6} \partial_m \partial_m \lambda + \hh \Big( \gamma -
      \frac{1}{2} \lambda' \Big) - \sigma \,,  & \qquad V_i &= \beta_i - \frac{1}{2}
      \epsilon_i' \,,  & \qquad h_{ij}^{TT} &= h_{ij}^{TT} \,, \\
      S_2 &= - \psi + \gamma' - \frac{1}{2} \lambda'' + \hh \Big( \gamma - \frac{1}{2}
      \lambda' \Big) - \sigma \,,  &  \rho_i &= \rho_i \,, \\
        S_3 &= \alpha - \sigma' \,.
    \end{aligned}
\end{equation}

\medskip\noindent
These equations are used to recast the Riemann tensor in the geodesic deviation equation
\eqref{Weylgeodesic1} in terms of gauge invariant observables:
\medskip
\begin{equation}
  \big[ \hat R^{i}{}_{00j} \big]^{(1)} = - \ddot S_1 \delta_{ij} + \partial_i \partial _j S_2
  + \frac{1}{2} \,\,\big( \partial_i \dot V_j + \partial_j \dot V_i \big) - \frac{1}{2} \,\ddot
  h_{ij}^{TT} - \partial_j \rho_i + \dot S_3 \delta_{ij} \,.
\end{equation}

\medskip
For comparison to Riemannian case of Einstein-Hilbert action in dS spacetime,
see eq.(\ref{EHdS}); notice the last two additional terms present here.

\subsection{Polarisation modes}
\label{sec_5.3: Polarisation modes}

We  choose to work in an enhanced version of the complete gauge fixing
gauge \eqref{GF1}:
\medskip
\begin{equation}
    \lambda = \gamma = \epsilon_i = \sigma = 0 \,. \label{GF2}
\end{equation}

\medskip\noindent
The first three conditions are identical to those in \eqref{GF1}, while the last one
ensures that $\lambda_D$ is fixed (this is unitary gauge for Weyl symmetry). 
At linear order, the equations of motion \eqref{WEOM} and \eqref{BEOM} become:
\medskip
\begin{align}
  0 &= - 2 \left( \nablabar_\nu \nablabar_\mu \hat R^{(1)} - 2 \nablabar_\nu \left( \omega_\mu
  \bar R \right) \right) + \left( 2 \bar R \hat R_{\mu\nu}^{(1)} + 2 \bar R_{\mu\nu} \hat R^{(1)}
  \right) - \Big( \bar g_{\mu\nu} \bar R \hat R^{(1)} + \frac{1}{2} \big( \delta g_{\mu\nu} \big) \bar
  R^2 \Big) \nonumber \\ 
  &\quad \, + \left( 2 \bar g_{\mu\nu} \nablabar^\alpha \nablabar_\alpha \hat R^{(1)} - 4 \bar
  g_{\mu\nu} \nablabar^\alpha \left( \omega_\alpha \bar R \right) \right)  \,, \label{linear_W} \\
  0 &= 3 d_1 \left[ \partial_\mu \hat R^{(1)} - 2 \omega_\mu \bar R \right] - d_2 \,
  \nablabar^\nu F_{\nu\mu} \label{linear_B}\,.
\end{align}
The expressions of linearised curvature terms in terms of gauge invariants are listed in Appendix~\ref{App_C}. Conformal time and spatial components of \eqref{linear_B} give:
\begin{align}
  0 &= 3 d_1 \partial_0 \hat R^{(1)} - 6 d_1 \bar R S_3 + d_2 a^{-2} \partial_m \partial_m S_3
  \,, \label{e2'} \\
  0 &= 3 d_1 \partial_i \hat R^{(1)} - 6 d_1 \bar R \rho_i - d_2 a^{-2} \left( \rho_i'' -
  \partial_m \partial_m \rho_i - \partial_i S_3' \right) \,, \label{e3'}
\end{align}
where $\rho_i' = \partial \rho_i / \partial \eta \equiv \partial_0 \rho_i$. From \eqref{linear_W} we find the (00), (0i) and (ij) equations:
\medskip
\begin{align}
  0 &= \bar R \left( - 2 \partial_m \partial_m S_2 + 6 S_1'' - 6 \hh S_2' + 6 \hh S_1' - 12
  \hh S_3 - 6 S_3' \right) - 2 \partial_m \partial_m \hat R^{(1)}  \nonumber \\ 
  &\quad+ 6 \hh \partial_0 \hat R^{(1)} + 6 \hh^2 \hat R^{(1)} - 12 \hh^2 \bar R S_2 \,,
  \label{e5'} \\
  0 &= - \partial_i \partial_0 \hat R^{(1)} + \hh \partial_i \hat R^{(1)} + \bar R \left( 2
  \partial_i S_1' + \frac{1}{2} \partial_m \partial_m V_i - 2 \hh \partial_i S_2 + 2 \hh V_i' +
  10 \hh^2 V_i \right) \,, \label{e6'} \\
  0 &= \delta_{ij} \left[ -2 \nablabar_\alpha \nablabar^\alpha \hat R^{(1)} + 4 \bar R
    \nablabar_\alpha \omega^\alpha + 2 \hh \partial_0 \hat R^{(1)} - 4 \hh \bar R S_3 + 6 \hh^2
    \hat R^{(1)} + \bar R \hat R^{(1)} \right] \nonumber \\
  &\quad - 2 \partial_j \partial_i \hat R^{(1)} + 4 \bar R \partial_j \rho_i + 2 \bar R \hat
  R_{ij}^{(1)} - \frac{1}{2} \left( a^2 h_{ij} \right) \bar R^2 \label{e7''} \,.
\end{align}
We can obtain an additional equation by taking the trace of \eqref{linear_W}:
\medskip
\begin{equation}
  \partial_0 \partial_0 \hat R^{(1)} + 2 \hh \partial_0 \hat R^{(1)} - \partial_m \partial_m
  \hat R^{(1)} - 2 \bar R \left( 2 \hh S_3 + S_3' \right) = 0 \,. \label{e9'}
\end{equation}

\medskip
Let us now investigate the sector of scalar fluctuations. If we contract $\partial_i$ into \eqref{e3'} and \eqref{e6'} we find
two additional identities which we use to simplify the remaining equations:
\begin{align}
    3 d_1 \hat R^{(1)} + d_2 a^{-2} S_3' &= 0 \,, \label{e4'} \\
    -\partial_0 \hat R^{(1)} + \hh \hat R^{(1)} + 2 \bar R S_1' - 2 \hh \bar R S_2 &= 0
    \,.\label{e7'}
\end{align}
We now use \eqref{e7'} together with \eqref{hatR1} to recast \eqref{e5'}:
\medskip
\begin{equation}
  \partial_m \partial_m \hat R^{(1)} = \bar R \left[ 2 \partial_m \partial_m S_1 - 9 \hh (S_1
    - S_2) - 9 \hh^2 (S_1 + S_2) \right] \,. \label{e10'}
\end{equation}

\medskip\noindent
We obtain one more identity by applying $\partial_0$ to \eqref{e7'}, and use this result together with \eqref{hatR1} to recast the trace equation \eqref{e9'} as follows:
\begin{equation}
  \partial_m \partial_m \hat R^{(1)} = \bar R \left[ \frac{2}{3} \partial_m \partial_m S_2 +
    \frac{4}{3} \partial_m \partial_m S_1 - 6 \hh (S_1 - S_2) - 6 \hh^2 (S_1 + S_2)
    \right] \,. \label{e11'}
\end{equation}

\medskip\noindent
This will be used shortly. Writing $\hat R_{ij}^{(1)}$ explicitly inside \eqref{e7''} using \eqref{Rij1}, contracting the
resulting expression with  $\partial_i \partial_j$, and using back the result into
\eqref{e7''} to substitute all the terms which were originally multiplied by $\delta_{ij}$,  we obtain:
\begin{multline}
  2 \delta_{ij} \left[ \partial_m \partial_m \hat R^{(1)} - \bar R \partial_m \partial_m (S_1 +
    S_2) \right] - 2 \partial_i \partial_j \hat R^{(1)} + 2 \bar R \partial_i \partial_j (S_1 +
  S_2) \\
  + \bar R \left( \partial_i V_j' + \partial_j V_i' \right) + 2 \bar R \hh \left( \partial_i V_j + \partial_j V_i \right) - \bar R \left( \partial_0 \partial_0 - \partial_m \partial_m + 2 \hh \partial_0 \right)
  h_{ij}^{TT} = 0 \,. \label{e}
\end{multline}
If we contract this equation by $\delta_{ij}$, we find the following identity:
\medskip
\begin{equation}
    \hat R^{(1)} = \bar R (S_1 + S_2) \,. \label{e12'}
\end{equation}
\medskip\noindent
Identity \eqref{e12'} can be further used to recast \eqref{e10'} and
\eqref{e11'}, respectively:
\begin{align}
    \bar R \left[ \partial_m \partial_m (S_1 - S_2) - 9 \hh (S_1 - S_2) - 9 \hh^2 (S_1 + S_2) \right] & = 0 \,,
  \label{e12''} \\
  \bar R \left[ \frac{1}{3} \partial_m \partial_m (S_1 - S_2) - 6 \hh (S_1 - S_2) - 6 \hh^2
  (S_1 + S_2) \right] &= 0 \,.
\end{align}
If we combine the last two identities above and use that $\bar R \neq 0$, we find:
\begin{equation}
    \partial_m \partial_m (S_2 - S_1) = 0 \quad \Rightarrow \quad S_2 = S_1 \,. \label{e13'}
\end{equation}

\medskip \noindent
If we use this inside \eqref{e12''} above, we find that $S_1 = S_2 = 0$. This in turn means that \eqref{e12'} becomes $\hat R^{(1)} = 0$, and,
consequently, \eqref{e4'} reduces to $S_3' = 0$. Using these results in \eqref{hatR1} we arrive at:
\begin{equation}
    12 a^{-2} \hh S_3 = 0 \quad \Rightarrow \quad S_3 = 0 \,.
\end{equation}

\medskip \noindent
Hence there are no scalar fluctuations present in the GW spectrum.

Having computed the scalar sector, we can now study the vector modes. Applying $\partial_j$ to \eqref{e7''}, and taking into account \eqref{Rij1} and the fact that all scalar invariants vanish, we arrive at:
\begin{equation}
    \partial_j \partial_j \left( V_i' + 2 \hh V_i \right) = 0 \quad \Rightarrow \quad V_i' + 2 \hh V_i = 0 \,. \label{Vj_eta}
\end{equation}

\medskip\noindent
This means that we can factor $V_i$ as $V_i(\eta,\textbf{x}) = \eta^2 B_i(\textbf{x})$, where $B_i(\textbf{x})$ is purely spatial. Next, setting all scalars to zero in \eqref{e6'}, we find:
\begin{equation}
  \bar R \left( \frac{1}{2} \partial_m \partial_m V_i + 2 \hh V_i' + 10 \hh^2 V_i \right) = 0
  \,,
\end{equation}

\medskip \noindent
which together with \eqref{Vj_eta} and previously outlined boundary conditions render $V_i = 0$. The presence of $\bar R$ as an overall factor highlights an important aspect of our analysis.
If we had chosen  a flat background from the very beginning (\ie $\bar R = 0$),
then all this information regarding the $V_i$ vector modes would have been missed. This is why de Sitter solutions are structurally different
from the flat case. Therefore, in a consistent analysis,
to see the full GW content in Weyl quadratic gravity,
one must first do all calculations in  a curved background, and only at
the very end may one take a suitable ``experimental''
flat space limit (\ie $\hh \rightarrow 0$),
consistent with current GW experiments. 

Since all scalar fluctuations vanish, from \eqref{e3'} we arrive at the following:
\medskip
\begin{equation}\label{TV}
  \left( \partial_0\partial_0 - \partial_m\partial_m \right) \rho_i - 72 \hh^2 \frac{d_1}{d_2}
  \rho_i = 0\,,
\end{equation}

\medskip \noindent
which is a modified wave equation for $\rho_i$. The last term is the mass ($m_\omega$) of the Weyl gauge field in $H_0$ units: $m_\omega^2/H_0^2 = - 72 d_1/d_2 = 12 \alpha_0^2 / \xi^2 > 0$, which agrees with the result for the mass in \cite{SMW} (for $q=2$).

For the remaining tensor modes one can simply set all scalar and vector invariants to zero inside \eqref{e}, which thus becomes identical to \eqref{wave_dS} up to an overall $\bar R$
factor:
\medskip
\begin{equation}\label{TT}
  \bar R \big( \partial_0 \partial_0 - \partial_m \partial_m + 2 \hh \partial_0 \big)
  h_{ij}^{TT} = 0 \,.
\end{equation}

\medskip\noindent
To summarise, the main results of this section are given by equation (\ref{TT}) for the graviton $h_{ij}^{TT}$ polarisation modes,
which generalises equation (\ref{wave_dS}),
and equations (\ref{TV}) for the two transverse vector polarisations $\rho_i$ ($\partial^i \rho_i = 0$) of
the Weyl gauge field, which are the only new modes present. No propagating scalar
modes are present.

\subsection{Comparison to experimental data}

The above analysis  shows that Weyl quadratic gravity  exhibits two additional transverse vector polarisation modes ($\rho_i$) on top of the two tensor modes ($h^{TT}_{ij}$) predicted
by Einstein-Hilbert gravity and recovered here.
Let us write  again  the equations of motion we found for the
corresponding (radiative) dof's in curved spacetime:
\begin{align}
  \Big( \partial_0\partial_0 - \partial_m\partial_m + \hh^2\, \frac{m_\omega^2}{H_0^2}
  \Big)\, \rho_i &= 0\,, \label{V_curved} \\[3pt]
  \left( \partial_0 \partial_0 - \partial_m \partial_m + 2\, \hh \,\partial_0 \right)
  h_{ij}^{TT} &= 0 \label{T_curved} \,.
\end{align}
\medskip
In the flat space limit (\ie $\hh \rightarrow 0$) consistent with the
GW experiment, then:
\begin{align}
  \left( \partial_0\partial_0 - \partial_m\partial_m \right) \rho_i &= 0\,,
  \label{V_flat} \\
  \left( \partial_0 \partial_0 - \partial_m \partial_m \right) h_{ij}^{TT} &= 0
  \label{T_flat} \,.
\end{align}
We discuss below these results in light of recent findings and explore potential
experimental imprints of the additional vector modes.

\paragraph{\bf $\bullet$ Geometric effect:\,\,}\mbox{}

\medskip\noindent
The current LVK network of detectors cannot resolve a GW onto all
six possible polarisations (one would require six independent unaligned detectors to do
that). Nonetheless it is possible to test the likelihood of various polarisation hypotheses against
each other.
The most recent analysis of the LVK collaboration \cite{LIGOScientific:2026oim}
tested six such scenarios against the purely tensorial (T) case, namely: Scalar (S), Vector
(V), Tensor-Scalar (TS), Tensor-Vector(TV), Vector-Scalar(VS) and Tensor-Vector-Scalar (TVS). Their results essentially rule out the S, V and VS scenarios (see Figure \ref{fig:LVK} below). The likelihoods of the remaining TS, TVS and TV scenarios have improved with the inclusion of recent experimental data, the largest increase being observed for the TV scenario (see Figure \ref{fig:LVK} below and Table 6 in \cite{LIGOScientific:2026oim}). Furthermore, in comparison to other scenarios, the TV scenario exhibits a steady increase in likelihood across the latest experimental studies performed by the LVK collaboration \cite{LIGOScientific:2026qni,LIGOScientific:2026oim}. Naively, this should not come as a surprise since
it is harder to accommodate a scalar, which is spatially isotropic, within a dominant
tensorial background, rather than a vector, which has a preferred direction, and thus can
(partially) align with one of the two directions of the tensor background, thereby improving the likelihood. However, one should not read too much into this, as the likelihoods of TS, TVS and TV scenarios are within experimental uncertainties.
\newline
\begin{figure}[h]
    \centering
    \includegraphics[width=0.68\textwidth]{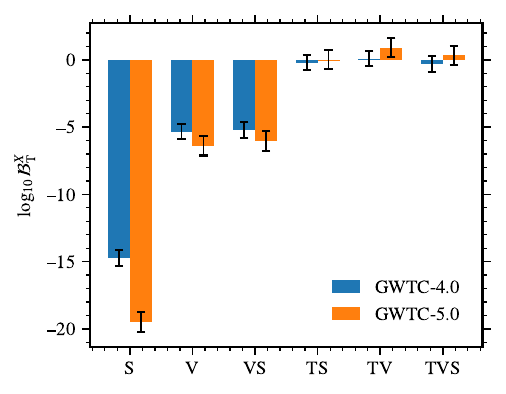}
    \caption{Comparison of likelihoods for different polarisation hypotheses
      against the purely tensor hypothesis. The left-hand columns (blue) represent previous results (GWTC-4.0), while the right-hand columns (orange) include the most recently analysed
      data (GWTC-5.0). In particular, note that in GWTC-5.0 the 90\% credible interval (\ie within error bars) for TV hypothesis excludes zero. Figure reproduced from \cite{LIGOScientific:2026oim} - Licensed under CC BY 4.0.}
    \label{fig:LVK}
\end{figure}

\paragraph{$\bullet$ Time-of-flight (TOF) and multi-messenger tests:\,\,} \mbox{}

\medskip\noindent
From \eqref{T_curved} one can infer that the two tensor modes $h_{ij}^{TT}$ behave
as plane waves which move at the speed of light on sub-Hubble scales (\ie $|k \eta| \gg 1$). The vector modes, on
the other hand, behave as massive modes, whose mass term in \eqref{V_curved},
\ie $\sqrt{72 \, d_1 / |d_2|} = m_\omega/H_0 \sim \alpha_0 M_p / H_0$, is coupled to curvature
$\hh \sim \eta^{-1}$. Strictly speaking, wave-like modes have $|k \eta| \gg m_\omega/H_0$, which
amounts to very short wavelengths for $\alpha_0 \sim O(1)$. Relaxing this condition,
for an ultra-weak Weyl gauge coupling \ie $\alpha_0 \ll 1$,
we find that vector modes can propagate slower than light, and thus slower
than the tensor modes. The dispersion relation, and the corresponding group
velocity, for each mode $k$ in \eqref{V_curved} read:
\begin{equation}
  \omega^2 = k^2 + \frac{m_\omega^2}{H_0^2 \eta^2} \qquad \Rightarrow \qquad v_g
  = \frac{d \omega}{d k} = \sqrt{1 - \left( \frac{m_\omega}{\omega H_0 \eta} \right)^2} \,.
  \label{dispersion}
\end{equation}
To measure a propagating wave coming from the massive vector modes, we require that
$\omega > m_\omega / (H_0 |\eta|)$, which means that $\omega \gtrsim \alpha_0 M_p H_0^{-1} |\eta|^{-1} \sim m_\omega$. Current GW detectors achieve their peak sensitivity around $f \approx 100$ Hz, which would correspond to a Weyl gauge boson mass $m_\omega \sim 10^{-13}$ eV. This is compatible with the limitations outlined below \eqref{mass_scales}, and, in addition to this, it is also consistent with dark matter analyses in which $\omega_\mu$ is a dark matter candidate \cite{Caputo:2021eaa,Cardoso:2018tly}. Furthermore, the fact that in a BBH merger each black hole has typically
around $20-30$ solar masses, exceeding $m_\omega$ by several orders of magnitude, indicates that it is possible in principle to detect imprints of $\omega_\mu$ within the GW spectrum. In fact, it has been argued that the GW 190521 event can be described as a merger between two vector boson stars (\ie Proca stars), corresponding to a vector boson mass of $8.7 \times 10^{-13}$ eV \cite{CalderonBustillo:2020fyi}.

In addition to this, the dispersion relation \eqref{dispersion} indicates a potential
spread of the chirp signal. One potential avenue to explore this effect is to search for GW events which have a gamma ray burst (GRB) counterpart, and compare the arrival time difference between GW and electromagnetic
wave signals. Current bounds for GW velocity $v_g$ are given in \cite{LIGOScientific:2017zic}:
\begin{equation}
    - \, 3 \, \times 10^{-15} < v_g - 1 < 7 \, \times 10^{-16} \,.
\end{equation}
Assuming vector modes are generated in this event, from the above bound and \eqref{dispersion}, we find that $\omega \sim 2 \pi \times 10^8 \times m_\omega$. For a mass $m_\omega \sim 10^{-13}$ eV, the characteristic frequency would be $f \sim 10^{-5}$ Hz which is outside the current range of detection for earth-based GW detectors ($\sim 10 - 10^4$ Hz). However, future detectors such as LISA whose designed detection range reaches the mHz region should be sensitive to $m_\omega \gtrsim 10^{-12} - 10^{-11}$ eV. While no particular mass is strongly favoured, larger values of $m_\omega$ remain viable and cannot be excluded.

\paragraph{$\bullet$ Post-Newtonian (PN) expansion:} \mbox{}\\
The PN framework describes the inspiral phase of the binary system as an
expansion in $v/c$, where $v$ is the orbital velocity. In Einstein-Hilbert gravity,
the leading contribution to the energy loss comes from the 0PN term which is a
quadrupole contribution. The presence of additional vector polarisation modes is
highlighted by extra energy leakage generated by a dipole term, which is encoded
by the -1PN coefficient. Most recent bounds on this can be found
in \cite{LIGOScientific:2026oim}.

\section{Conclusions}
\label{sec_6: Conclusions}

Weyl conformal geometry with its associated action is a (\textit{vector-tensor}) gauge theory of the
Weyl group of dilatations and Poincar\'e symmetry. The Weyl quadratic (gauge theory
of) gravity action (WQG) and its generalisation, the WDBI action, are Weyl-anomaly
free candidates for a quantum gauge theory of gravity. They exhibit a spontaneous Weyl gauge symmetry
breaking in which the Weyl gauge boson of dilatations becomes massive, decouples,
and one recovers in the broken phase the  Einstein-Hilbert action plus a
positive cosmological constant.
In order to make contact with experimental tests, we studied the GW that such
theory generates. 

Let us briefly summarise  our results. In section~\ref{sec_3.2: Weyl geometry}
we derived the geodesic deviation equation in Weyl conformal geometry. Key
to this was the implementation of the Weyl gauge covariant formulation reviewed in
section~\ref{sec_2: Weyl geometry...}, which, being metric-compatible, allowed us
to carry on the arguments from Riemannian geometry to Weyl geometry.

The polarisation modes for WQG action \eqref{S} were computed in
section~\ref{sec_5: Weyl quadratic gravity}. Here, besides the topological term (which is a total
derivative in $d = 4$ dimensions), we did not include the Weyl-tensor-squared term (or equivalently
$\hat R_{\mu\nu} \hat R^{\mu\nu}$) from \eqref{full_Weyl_action} into our analysis. Since this is
identical to its Riemannian counterpart, one can show that, with appropriate boundary
conditions imposed on the metric, the behaviour of this term is equivalent to the Einstein-dS
case \cite{Maldacena:2011mk}. Thus, we expect the dynamical content of the theory to not be
modified by this. The ultimate case of GW in the strong field regime would be to consider the case
of the WDBI action, however this case becomes much more involved due to its
high non-linearity \cite{InPrep}.

The gauge invariant variables are constructed in section~\ref{sec_5.2: SVT decomposition}, and
in section~\ref{sec_5.3: Polarisation modes} we used them to recast the equations of motion.
Consequently, we showed that all scalar gauge invariants vanish. In addition to the two
transverse-traceless tensor modes predicted by EH gravity,
and recovered here, we found two additional
transverse vector modes, induced by the fluctuations of the Weyl gauge field of
dilatations $\omega_\mu$. In the process, we found that both vector and tensor
elementary dof's of the metric fluctuations
couple to the background  \ie they appear multiplied by an overall factor of $\bar R$. This
means that, had one used a flat background to begin with, crucial information would have been
lost, and this in turn highlights the importance of our approach. 

Compared to other scenarios, the \textit{vector-tensor} scenario appears to provide the most promising direction for investigations of gravity theories beyond Einstein-Hilbert gravity. In particular, the GW 190521 as interpreted by Bustillo et al. could be direct evidence of vector polarisation modes, and thus of Weyl gauge symmetry. Our results can be used in searches for new (transverse) vector polarisation modes in the GW spectrum. If detected, they will be important evidence for  Weyl gauge symmetry and for an underlying Weyl gauge theory beyond Einstein-Hilbert gravity.

\newpage
\section*{Appendix}

\def\theequation{A-\arabic{equation}}
\def\thesection{A}
\setcounter{equation}{0}

\section{Weyl geometry curvatures}
\label{App_A}

We outline below the identities relating Weyl curvatures to
their Riemannian counterparts in $d=4$ dimensions, in the
Weyl gauge covariant (metric!) formulation of Weyl geometry,
for details see \cite{DG1,CDA2}:
\begin{align}\label{a1}
  \hat R_{\rho\sigma\mu\nu} &= R_{\rho\sigma\mu\nu} + \left[ g_{\mu\sigma} \nabla_\nu \omega_\rho -
    g_{\mu\rho} \nabla_\nu \omega_\sigma + g_{\nu\rho} \nabla_\mu \omega_\sigma - g_{\nu\sigma}
    \nabla_\mu \omega_\rho \right] \nonumber \\[5pt]
  &+ \omega^2 \left( g_{\mu\sigma} g_{\nu\rho} - g_{\mu\rho} g_{\nu\sigma} \right) + \omega_\mu \left(
  \omega_\rho g_{\nu\sigma} - \omega_\sigma g_{\nu\rho} \right) + \omega_\nu \left( \omega_\sigma
  g_{\mu\rho} - \omega_\rho g_{\mu\sigma} \right) \,, \\[5pt]
  \hat R_{\mu\nu} &= R_{\mu\nu} - 2 \nabla_\nu \omega_\mu - g_{\mu\nu} \nabla_\alpha \omega^\alpha +
  2 \left( \omega_\mu \omega_\nu - g_{\mu\nu} \omega_\alpha \omega^\alpha \right) \,, \\[5pt]
    \hat R &= R - 6 \nabla_\mu \omega^\mu - 6 \omega_\mu \omega^\mu \,, \\[5pt]
    \hat F_{\mu\nu} &= F_{\mu\nu} = \partial_\mu \omega_\nu - \partial_\mu \omega_\nu \,.
\end{align}

\medskip\noindent
where all operators  in the right hand side are those of Riemannian geometry. 

The Chern-Euler-Gauss-Bonnet term in Weyl geometry is defined as follows:
\begin{equation}
    \hat G = \hat R^2 - 4 \hat R_{\mu\nu} \hat R^{\nu\mu} + \hat R_{\mu\nu\rho\sigma} \hat R^{\rho\sigma\mu\nu} \,.
\end{equation}
In $d=4$ dimensions considered in this work, $\hat G$ does not affect the equations of motion.

\def\theequation{B-\arabic{equation}}
\def\thesection{B}
\setcounter{equation}{0}

\section{Commutator of coordinate vectors in Weyl geometry}
\label{App_B}

Here we show explicitly that eq.\eqref{commutatorWeyl} in the text holds true.
First, we can recast this equation by taking into account that the spacetime Weyl
charge of a coordinate vector field is $\tilde q = -1$, then:  
\medskip
\begin{align}
  X^\mu \Big(\partial_\mu S^\lambda + \tilde \Gamma^\lambda_{\mu\nu} S^\nu - \omega_\mu 
  S^\lambda \Big) &= S^\mu \Big(
  \partial_\mu X^\lambda + \tilde \Gamma^\lambda_{\mu\nu}  X^\nu - \omega_\mu
   X^\lambda\Big) \\[4pt]
  \Rightarrow X^\mu \partial_\mu S^\lambda - \omega_\mu X^\mu S^\lambda &= S^\mu \partial_\mu
  X^\lambda - \omega_\mu S^\mu X^\lambda \,, \label{b2}
\end{align}

\medskip\noindent
where in going to the second line we have used the fact that $\tilde \Gamma$ is symmetric in
its lower indices. Using the definitions of $X^\mu$ and $S^\mu$ in \eqref{tangentvector} and
\eqref{deviationvector}, respectively, we can rewrite the first terms on both sides as
follows:
\medskip
\begin{equation}
  X^\mu \partial_\mu S^\lambda = \frac{\partial S^\lambda}{\partial \tau} = \frac{\partial^2
    x^\lambda}{\partial \tau \partial s} \,, \qquad S^\mu \partial_\mu X^\lambda = \frac{\partial
    X^\lambda}{\partial s} = \frac{\partial^2 x^\lambda}{\partial s \partial \lambda} \,.
  \label{b3}
\end{equation}

\medskip\noindent
Given that partial derivatives commute, the above expressions are identical,
then proving \eqref{b2} is equivalent to showing that:
\medskip
\begin{equation}
    \omega_\mu \left( X^\mu S^\lambda - S^\mu X^\lambda \right) = 0 \,. \label{b2a}
\end{equation}

\medskip\noindent
To show this, we first note that the spacetime Weyl charge of $X^\mu = \partial x^\mu / \partial \tau$ (or $S^\mu = \partial x^\mu / \partial s$) is the same as the Weyl charge of the corresponding differential operator \cite{Hobson_2020} which acts on coordinates since the
coordinates $x^\mu$ remain unchanged under Weyl gauge transformations. This observation
allows us to perform a Weyl gauge transformation on the two terms from \eqref{b3}:
\medskip
\begin{align}
  \frac{\partial S^\mu}{\partial \tau} = \frac{\partial}{\partial \tau} \frac{\partial
    x^\mu}{\partial s} \rightarrow \Sigma^{-1} \frac{\partial}{\partial \tau} \left(
  \Sigma^{-1} \frac{\partial x^\mu}{\partial s} \right) = \Sigma^{-2} \left( \frac{\partial^2
    x^\mu}{\partial \tau \partial s} - S^\mu X^\alpha \partial_\alpha \ln \Sigma \right) \,,
  \label{b5} \\
  \frac{\partial X^\mu}{\partial s} = \frac{\partial}{\partial s} \frac{\partial
    x^\mu}{\partial \tau} \rightarrow \Sigma^{-1} \frac{\partial}{\partial s} \left(
  \Sigma^{-1} \frac{\partial x^\mu}{\partial \tau} \right) = \Sigma^{-2} \left( \frac{\partial^2
    x^\mu}{\partial s \partial \tau} - X^\mu S^\alpha \partial_\alpha \ln \Sigma \right) \,,
  \label{b6}
\end{align}

\medskip\noindent
where we also use the fact that $\left( \partial / \partial \tau \right) \Sigma(x) = X^\alpha
\partial_\alpha \Sigma(x)$, and similarly for $\partial / \partial s$. The LHS of \eqref{b5}
and \eqref{b6} are identical, thus the RHS must coincide as well:
\medskip
\begin{equation}
  \left( S^\mu X^\alpha - X^\mu S^\alpha \right) \partial_\alpha \ln \Sigma = 0, \quad \forall \,
  \Sigma = \Sigma(x) \,.
\end{equation}
This means that $S^\mu X^\alpha = S^\alpha X^\mu$, and, if we contract this with $\omega_\mu$, we
obtain precisely \eqref{b2a}.

\def\theequation{C-\arabic{equation}}
\def\thesection{C}
\setcounter{equation}{0}

\section{Linear level expressions}
\label{App_C}
We outline below the expressions used in section~\ref{sec_5.3: Polarisation modes},
for the various Weyl curvature terms as functions of gauge invariant quantities in the gauge
chosen in eq.\eqref{GF2}, and at linear order in fluctuations. 
\begin{itemize}
    \item The Weyl-Ricci tensor:
        \begin{align}
          \hat R_{00}^{(1)} &=  - \partial_i \partial_i S_2 + 3 S_1'' - 3 \hh S_2' + 3 \hh
          S_1' - 4 \hh S_3 - 3 S_3' \,, \\
          \hat R_{0i}^{(1)} &= 2 \partial_i S_1' + \frac{1}{2} \partial_j \partial_j V_i - 2
          \hh \partial_i S_2 + 2 \hh V_i' + 7 \hh^2 V_i - 2 \partial_i S_3 + 2 \hh \rho_i \,, \\
          \hat R_{i0}^{(1)} &= 2 \partial_i S_1' + \frac{1}{2} \partial_j \partial_j V_i - 2
          \hh \partial_i S_2 + 2 \hh V_i' + 7 \hh^2 V_i - 2 \rho_i' + 2 \hh \rho_i \,, \\
          \hat R_{ij}^{(1)} &= \delta_{ij} \left[ - S_1'' + \partial_m \partial_m S_1 + 6 \hh
            (S_2 - S_1) - 5 \hh S_1' + \hh S_2' \right] \nonumber \\
          &+ \partial_i \partial_j (S_1 + S_2) - \frac{1}{2} \left( \partial_0 \partial_0 -
          \partial_m \partial_m \right) h_{ij}^{TT} + \frac{1}{2} \left( \partial_i V_j' +
          \partial_j V_i' \right) \label{Rij1}\\
          &+ \hh \left( \partial_i V_j + \partial_j V_i \right) - 3 \hh^2 h_{ij}^{TT} - \hh
          \partial_0 h_{ij}^{TT} - 2 \partial_j \rho_i + 4 \hh S_3 \delta_{ij} + S_3' \delta_{ij}
          \,. \nonumber
        \end{align}

    \item The Weyl-Ricci scalar:
        \begin{align}
          \hat R^{(1)} = a^{-2} \big[ &- 2 \partial_m \partial_m S_2 - 4 \partial_m \partial_m
            S_1 + 6 S_1'' \nonumber \\
            &+ 18 \hh S_1' - 6 \hh S_2' + 18 \hh (S_1 - S_2) - 6 \hh^2 ( - 3 S_1 + S_2)
            \label{hatR1}\\
            &- 12 \hh S_3 - 6 S_3' \big] \,. \nonumber
        \end{align}

    \item The gauge kinetic term $F_{\mu\nu}$:
        \begin{align}
            F_{0i} &= \rho_i' - \partial_i \alpha \,, \\
            F_{ij} &= \partial_i \rho_j - \partial_j \rho_i \,.
        \end{align}
\end{itemize}
The following identites are useful throughout the paper for the various computations that we perform:
\medskip
\begin{align}
    \nablabar_\alpha \omega^\alpha &= a^{-2} \left( 2 \hh S_3 + S_3' \right) \,, \\
    \nablabar_\mu \omega_\nu &= 
    \begin{cases}
        \nablabar_0 \omega_0 = \hh S_3 + S_3' \,, \\
        \nablabar_i \omega_0 = \partial_i S_3 - \hh \rho_i \,, \\
        \nablabar_0 \omega_i = \rho_i' - \hh \rho_i \,, \\
        \nablabar_i \omega_j = \partial_i \rho_j - \hh S_3 \delta_{ij} \,,
    \end{cases} \\
    \nablabar_\mu \nablabar_\nu \hat R^{(1)} &= 
    \begin{cases}
        \nablabar_0 \nablabar_0 \hat R^{(1)} = \partial_0 \partial_0 \hat R^{(1)} - \hh \partial_0 \hat R^{(1)} \,, \\
        \nablabar_0 \nablabar_i \hat R^{(1)} = \partial_0 \partial_i \hat R^{(1)} - \hh \partial_i \hat R^{(1)} \,, \\
        \nablabar_i \nablabar_0 \hat R^{(1)} = \partial_i \partial_0 \hat R^{(1)} - \hh \partial_i \hat R^{(1)} \,, \\
        \nablabar_i \nablabar_j \hat R^{(1)} = \partial_i \partial_j \hat R^{(1)} - \hh \delta_{ij} \partial_0\hat R^{(1)} \,,
    \end{cases} \\
    \nablabar_\alpha \nablabar^\alpha \hat R^{(1)} &= a^{-2} \left( \partial_0 \partial_0 - \partial_i \partial_i + 2 \hh \partial_0 \right) \hat R^{(1)} \,.
\end{align}

\bibliographystyle{hunsrt}
\bibliography{references_V2}
\end{document}